\documentclass[12pt]{article}
\usepackage{amsmath,amsthm,amsfonts, latexsym}
\usepackage{graphicx}

\begin{document}

\begin{center}

\bigskip

{{\LARGE Power of unentangled measurements on two antiparallel
spins\\}

\bigskip
{\large T. V\'ertesi\\} Institute of Nuclear Research of the Hungarian
Academy of Sciences, H-4001 Debrecen, P.O. Box 51, Hungary
\\tvertesi@dtp.atomki.hu
} \vspace{3cm}

\end{center}



\begin{abstract}
We consider a pair of antiparallel spins polarized in a random direction to
encode quantum information. We wish to extract as much information as
possible on the polarization direction attainable by an unentangled
measurement, i.e., by a measurement, whose outcomes are associated with
product states. We develop analytically the upper bound $0.7935$ bits to the
Shannon mutual information obtainable by an unentangled measurement, which
is definitely less than the value $0.8664$ bits attained by an entangled
measurement. This proves our main result, that not every ensemble of product
states can be optimally distinguished by an unentangled measurement, if the
measure of distinguishability is defined in the sense of Shannon. We also
present results from numerical calculations and discuss briefly the case of
parallel spins.
\end{abstract}

\vfill

\vfill
\newpage

\section{Introduction}

One of the central problems in quantum information theory is the state
discrimination problem. Suppose that one is given a single quantum system,
which is known to be in one of several possible states with a certain
\textit{a priori} probability.  Then one wishes to carry out such a
measurement on the system that would yield as much information about the
identity of the system's state as possible, where the gained information is
defined in terms of the Shannon mutual information.

Although there exist other figures of merit, which quantify
distinguishability, such as the statistical overlap (i.e., the fidelity), or
the Kullback-Leibler relative information \cite{Helstrom,Fuchs}, in this
work we will focus on the mutual information, which quantifies the quality
of measurement through the average gain of information about the unknown
states \cite{Peres,Shor}.

A particular instance of the discrimination problem is when each possible
state of the system is restricted to be in a product state. With regard to
this, some time ago Peres and Wootters \cite{PW} addressed the intriguing
problem of whether in order to gain as much information as possible from an
ensemble of product states it is sufficient to do local measurements or
sometimes necessary to carry out a global measurement on the system as a
whole. Technically, in the first case one is permitted to do any sequence of
local operations carried out on each subsystem individually and classical
communication between the subsystems (LOCC), while in the second case
arbitrary quantum operations are allowed on both spins.

Hitting upon a special ensemble of states, the double-trine states, they
showed evidence that a global measurement was distinctly better than any
LOCC measurement. Recently, Decker \cite{Decker} confirmed this result and
other studies \cite{Bennett1,Bennett2,DiVincenzo} also proved conclusively
the superiority of global measurements over LOCC measurements, for which
property the phrase ``quantum nonlocality without entanglement'' was coined
\cite{Bennett1}.

While in the above case, a distinction was made between the power of global
and local measurements, one may further divide global measurements into the
following two distinct classes: {\textit{Unentangled}} measurements, whose
outcomes are associated with product states, and \textit{entangled}
measurements, for which at least one outcome is associated with an entangled
state. An interesting question was posed recently by Wootters
\cite{Wootters} of whether every ensemble of product states could be
distinguished just as well by an unentangled measurement as by an entangled
measurement. Although it turned out \cite{Wootters}, that an unentangled
measurement on the double-trine ensemble was as good as an entangled
measurement, the question remained open about the existence of other kinds
of product states where the best unentangled measurement could be beaten by
an entangled one.

In the present article we wish to address this general question by focusing
on the following special state discrimination problem: Given a source, which
emits a pair of antiparallel spin-$1/2$ particles (spins for short)
polarized along a random space direction, the observer's task is to
perform an unentangled measurement on the two spins which provides the
maximum gain of information about the polarization direction. In the present
study we manage to bound from above the maximum gain of information
attainable by an unentangled measurement on two antiparallel spins, and this
upper bound will appear to be smaller than the information which can be
extracted by a particular entangled measurement. With this result we intend
to give an answer for the question raised above, that on product states
entangled measurements are in general more informative than unentangled
ones. Further, since the set of unentangled measurements is strictly larger
than the set of LOCC measurements \cite{Bennett1}, the pair of antiparallel
spins can be considered as another example beside the double-trine
ensemble, where global measurements are distinctly more powerful than LOCC
measurements.

Note that the state discrimination of antiparallel spins discussed above
can be interpreted as a quantum communication problem, i.e., the problem of
communicating an unknown spatial direction between two distant parties by
the transmission of quantum particles. This problem has been extensively
studied in the literature \cite{GP,Massar,PS,Bagan1,BM,Jeffrey,Bartlett1},
but using the fidelity as a figure of merit. Our findings corresponding to
the mutual information thus can also be regarded as a complement to the
results of the cited references.

The article is organized as follows: In Sec.~2 we introduce the notation and
formulate the problem. In Sec.~3 the rotational invariance property of the
mutual information is demonstrated and the problem of obtaining the best
unentangled measurement is presented as a constrained optimization problem.
In Sec.~4 the optimization is performed by the Lagrange multipliers method
by applying Jensen's inequality. Then the best unentangled measurement is
given explicitly in terms of measurement projectors and we also discuss
briefly the case of two parallel spins. The paper concludes in Sec.~5 with a
discussion of the results.

\section{Formalism}
\subsection{POVM measurement}

As we mentioned in the Introduction our state discrimination problem can be
presented as a quantum communication task: Suppose Alice wishes to
communicate to Bob a spatial direction, i.e., a unit vector $\mathbf n$
chosen completely at random. In order to accomplish the task, Alice prepares
two spins in the product state
\begin{equation}
|A(\mathbf n)\rangle=|\mathbf n\rangle|-\mathbf n\rangle \;, \label{signal}
\end{equation}
where the first spin is polarized along the random space direction $\mathbf
n$ and the second spin is polarized in the opposite direction $-\mathbf n$.
Then she sends the pair of antiparallel spins to Bob, and upon receiving it
Bob performs an unentangled measurement on the spins so as to acquire as
much knowledge about the spatial direction $\mathbf n$ as possible. The
polarized spin state $|\mathbf n\rangle$ corresponding to Alice's signal
satisfies
\begin{equation}
\hat{\sigma} \cdot \mathbf n|\mathbf n\rangle=|\mathbf n\rangle \;,
\label{eigeq}
\end{equation}
where $\hat{\sigma}$ are the usual Pauli matrices.

On the other hand, the mathematical representation of Bob's measurement apparatus
is a positive operator valued measure (POVM) consisting of a set of operators
$E_r$, which sum up to unity on the four-dimensional Hilbert space of the
two spins,
\begin{equation}
\sum_{r=1}^M{E_r= \mathbb I}\;, \label{POVM}
\end{equation}
where $r = 1, \ldots, M$ labels the outcome of the measuring process and we
require $M\geq4$ owing to the size of the Hilbert space. Note that the sum
in Eq.~(\ref{POVM}) can be extended to the continuous case as well by a
suitable adjustment of the notation. Taking into account that one can always
assume the projectors $E_r$ to have rank one \cite{Davies}, we can write
\begin{equation}
E_r=c_r|\psi_r\rangle\langle\psi_r| \;, \label{projector}
\end{equation}
where $c_r$ are positive weights and states $|\psi_r\rangle$ are normalized.
Bob is allowed to carry out unentangled measurements, i.e., measurements for
which each of the POVM operator elements $E_r$ is a tensor product. Thus
each state $|\psi_r\rangle$ corresponding to measurement outcome $r$ can be
written in the product form
\begin{equation}
|\psi_r\rangle=|\mathbf{n}_{1r}\rangle|\mathbf{n}_{2r}\rangle\;.
\label{prodr}
\end{equation}
The pairs of unit vectors $\mathbf{n}_{1r}$ and $\mathbf{n}_{2r}$ are yet
free parameters, which must be adjusted by Bob appropriately so as to
achieve the highest possible amount of mutual information between the
outcomes of his unentangled measurement and Alice's states. In order to
arrive at an explicit formula for this information gain let us introduce
some notations.

\subsection{Information gain}
The conditional probability $p(r|\mathbf n)$ that Alice's preparation
$|A(\mathbf n)\rangle$ yields Bob's result $r$ is given by Born's rule
\begin{equation}
p(r|\mathbf{n})=c_r|\langle A(\mathbf{n})|\psi_r\rangle|^2\;, \label{Born}
\end{equation}
which on substitution the signal state~(\ref{signal}) and Bob's product
states~(\ref{prodr}) into this expression gives
\begin{equation}
p(r|\mathbf{n})=c_r|\langle \mathbf n|\mathbf n_{1r}\rangle|^2 |\langle
-\mathbf n|\mathbf n_{2r}\rangle|^2        \;. \label{prn}
\end{equation}

Let us designate an arbitrary point $(\theta,\phi)$ on the Bloch sphere by
the unit vector $\mathbf n(\theta,\phi)$ specified by the coordinates
$\mathbf n=(\cos\phi\sin\theta,\sin\phi\sin\theta,\cos\theta)$. Since Alice
chooses $\mathbf n$ randomly, or say equivalently, Bob has no knowledge
before his measurement about the space direction $\mathbf n$ which Alice
indicates by her signal~(\ref{signal}), it entails the uniform {\textit{a
priori}} distribution $p(\mathbf n)=1$ on the Bloch sphere.

The {\textit{a priori}} probability that Bob has measurement outcome $r$ is
\begin{equation}
p(r)=\int{dn p(r|\mathbf n)p(\mathbf n)}\;, \label{pr}
\end{equation}
where the integration is performed over the whole Bloch sphere and
$dn=\frac{1}{4\pi}\sin\theta d\theta d\phi$ is the uniform measure on the
Bloch sphere. Then applying Bayes' theorem the {\textit{a posteriori}}
probability for $\mathbf n$ is given by
\begin{equation}
p(\mathbf n|r)=\frac{p(r|\mathbf n)}{p(r)}p(\mathbf n). \label{pnr}
\end{equation}

The Shannon mutual information is the average amount of information that one
gains about the direction $\mathbf n$ upon observing the outcome of the
measurement. Thus it can be written as the difference of the {\textit{a
priori}} entropy $H_{initial}$ of $p(\mathbf n)$ and the average {\textit{a
posteriori}} entropy $\bar{H}_{final}$ of $p(\mathbf n|r)$
\cite{Peres,Shor}.

The value of $H_{initial}$ is infinite for the continuous distribution
$p(\mathbf n)$, but as it can be shown \cite{Peres,Bagan2} the divergent
term is cancelled by terms from $\bar{H}_{final}$ and the Shannon mutual
information can be expressed in the closed form \cite{TV,ATV,Bartlett2}
\begin{equation}
I_{av}=\sum_{r=1}^M{p(r)K\left(p(\mathbf n|r)/p(\mathbf n)\right)}\;,
\label{mutual}
\end{equation}
in terms of the Kullback-Leibler relative information between the
distributions $p(\mathbf n|r)$ and $p(\mathbf n)$,
\begin{equation}
K\left(p(\mathbf n|r)/p(\mathbf n)\right)=\int {dn p(\mathbf n|r)\log_2
\frac{p(\mathbf n|r)}{p(\mathbf n)}} \;. \label{Kullback}
\end{equation}

Our starting point is this information gain, Eq.~(\ref{mutual}), to quantify
Bob's measuring strategy, which is well-defined for continuous distributions
\cite{Hobson}. Particularly, we intend to optimize Eq.~(\ref{mutual}) by
restricting Bob to perform an unentangled measurement described by
Eq.~(\ref{prodr}) and considering that the \textit{a priori} distribution of
Alice's ensemble is $p(\mathbf n)=1$. However, we also want the
projectors $E_r$ to constitute a valid POVM. This imposes the following pair of
constraints, which Bob's unentangled measurement operators must fulfill in
order to optimize his gained information~(\ref{mutual}),
\begin{equation}
\sum_{r=1}^M{c_r} = 4\;, \hspace{7mm} \sum_{r=1}^M{p(r)} =1  \; ,
\label{constraints}
\end{equation}
where the first constraint is obtained by evaluating the trace of the POVM
condition~(\ref{POVM}) considering Eq.~(\ref{projector}), and the second
constraint is due to the fact that $p(r)$ is a probability distribution.

\section{Optimization problem}
\subsection{Rotational invariance}

As a next step, we aim to exploit rotational invariance properties of the
{\textit{a priori}} probability $p(r)$ defined by Eq.~(\ref{pr}) and the
mutual information $I_{av}$ given by Eq.~(\ref{mutual}) in order that we
could bring the state~(\ref{prodr}) to a simpler form. Regarding the
uniform distribution $p(\mathbf n)=1$ and substituting formula~(\ref{prn})
into the definition~(\ref{pr}) one obtains
\begin{equation}
p(r)=c_r\int{dn |\langle \mathbf n|\mathbf n_{1r}\rangle|^2 |\langle
-\mathbf n|\mathbf n_{2r}\rangle|^2  } \;.
\end{equation}
This formula, owing to the rotational invariance of the integral, is
unchanged under an arbitrary collective rotation $R_r$ of the unit vectors
$\mathbf n_{1r}$ and $\mathbf n_{2r}$, i.e.,
\begin{equation}
p(r)=c_r\int{dn |\langle \mathbf n|R_r(\mathbf n_{1r})\rangle|^2 |\langle
-\mathbf n|R_r(\mathbf n_{2r})\rangle|^2  } \;.
\end{equation}
For the same symmetry reasons the information gain~(\ref{mutual}) (with
$p(\mathbf n)=1$) also remains invariant by replacing $\mathbf
n_{ir}\rightarrow R_r(\mathbf n_{ir}),\hspace{2.5mm} i=1,2$. In particular,
let us choose the rotations $R_r$ in such a way that
\begin{align}
   R_r(\mathbf n_{1r})&=\mathbf z \;,\nonumber \\
   R_r(\mathbf n_{2r})&=\mathbf n_r(\theta_r,\phi_r=0)
  \label{rotate}
\end{align}
for r = 1, \ldots, M. That is, by a suitable collective rotation of the pair
of unit vectors $\mathbf n_{1r}$ and $\mathbf n_{2r}$, one rotates $\mathbf
n_{1r}$ into the north pole, while $\mathbf n_{2r}$ to a point, represented
by $\mathbf n_r$, so that it lies on the polar great circle arc of the Bloch
sphere. Since $R_r$ represents an arbitrary rotation, the
rotations~(\ref{rotate}) can always be performed, also guaranteeing
\begin{equation}
\mathbf z\cdot \mathbf n_r=\mathbf n_{1r}\cdot \mathbf
n_{2r}=\cos\theta_r\;, \label{angle}
\end{equation}
where $\theta_r$ is the angle between the pair of vectors $\mathbf n_{1r}$
and $\mathbf n_{2r}$. Therefore, in effect the mapping of states
\begin{equation}
|\psi_r\rangle=|\mathbf n_{1r}\rangle|\mathbf
n_{2r}\rangle\rightarrow|\tilde{\psi}_r\rangle=|\mathbf z\rangle|\mathbf
n_r\rangle \label{map}
\end{equation}
induced by the collective rotations $R_r$ will not change the amount of
information gain~(\ref{mutual}). Here $|\mathbf n_r\rangle$ can be written
explicitly in the basis $\{|\mathbf z\rangle,|-\mathbf z\rangle\}$
\begin{equation}
|\mathbf n_r\rangle=\cos\frac{\theta_r}{2}|\mathbf
z\rangle+\sin\frac{\theta_r}{2}|-\mathbf z\rangle \label{nr}
\end{equation}
using relation~(\ref{angle}).

\subsection{Constrained formula}
Let us make use of the informational equivalence which we found in the
preceding subsection between $|\psi_r\rangle$ and $|\tilde{\psi}_r\rangle$
and make the replacement~(\ref{map}) for obtaining a simplified form of the
information gain~(\ref{mutual}). Then by means of Eq.~(\ref{map}) and
considering $p(\mathbf n)=1$ the conditional probability $p(r|\mathbf n)$ in
Eq.~(\ref{prn}) is mapped to
\begin{align}
p(c_r,\theta_r|\mathbf n) &=c_r |\langle \mathbf n|\mathbf z\rangle|^2
|\langle
-\mathbf n|\mathbf n_r\rangle|^2 \nonumber \\
&=\frac{c_r}{2}\cos^2\frac{\theta}{2}(1-\cos\theta\cos\theta_r-\cos\phi\sin\theta\sin\theta_r)
\label{mapprn}
\end{align}
and in turn the probability $p(r)$  becomes
\begin{equation}
p(c_r,\theta_r)=\int{dn p(c_r,\theta_r|\mathbf
n)}=c_r\frac{3-\cos\theta_r}{12}\;, \label{mappr}
\end{equation}
where in Eq.~(\ref{mapprn}) the variables $(c_r,\theta_r,\theta)$  are
written out explicitly and were also used in the evaluation of
Eq.~(\ref{mappr}).

Given Eqs.~(\ref{mapprn}-\ref{mappr}) the problem of optimizing Bob's
information gain~(\ref{mutual}) subject to the corresponding
constraints~(\ref{constraints}) can be presented in terms of the variables
$(c_r,\theta_r), \hspace{2.5mm} r = 1, \ldots, M$. Namely, after a bit of
algebra and using $p(\mathbf n)=1$ the information gain~(\ref{mutual})
quantified by the mutual information takes the explicit form
\begin{equation}
I_{av}=\sum_{r=1}^M {c_r I(\theta_r)}\;, \label{Iav}
\end{equation}
where
\begin{equation}
I(\vartheta)=\frac{3-\cos\vartheta}{12}\int{dn \frac{p(c_r,\vartheta|\mathbf
n)}{p(c_r,\vartheta)}\log_2{\frac{p(c_r,\vartheta|\mathbf
n)}{p(c_r,\vartheta)}} }\;. \label{I}
\end{equation}
Note that as a consequence of Eqs.~(\ref{mapprn}) and (\ref{mappr}) the
fraction $p(c_r,\vartheta|\mathbf n)/p(c_r,\vartheta)$ and hence
$I(\vartheta)$ within Eq.~(\ref{I}) are independent of the index $r$.

Since the information gain~(\ref{Iav}) is subjected to constraints we have
to impose some restrictions on the domain of the variables $(c_r,\theta_r)$.
On the one hand, these variables need to be in the range
\begin{equation}
(c_r>0,\;0\leq\theta_r\leq \pi),\hspace{5mm} r = 1, \ldots, M \;,
\label{space}
\end{equation}
where the number $M$ is at least $4$. On the other, the
constraints~(\ref{constraints}) further restrict the domain and these
conditions can be brought to the explicit forms
\begin{equation}
\sum_{r=1}^M{c_r} = 4\;, \hspace{7mm} \sum_{r=1}^M{c_r\cos\theta_r} =1  \; ,
\label{newconstraints}
\end{equation}
by replacing Eq.~(\ref{pr}) with Eq.~(\ref{mappr}). In the following, let us
refer to the domain, which is within the range~(\ref{space}) and satisfies
constraints~(\ref{newconstraints}) as the feasible region.

To summarize this section, the information gain in Eq.~(\ref{Iav}) with the
constraints~(\ref{space}-\ref{newconstraints}) constitute the constrained
optimization problem: Bob's task is to maximize the mutual information
(defined by Eq.~(\ref{Iav})) between his unentangled measurement outcomes
and Alice's signals by choosing appropriately the set of values
${(c_r,\theta_r)}$ from the feasible region (defined by
Eqs.~(\ref{space}-\ref{newconstraints})). The next section is centered on
the problem of how to build a reasonable upper bound to this maximal amount of
information gain.

\section{Solution}
\subsection{Upper bound}

The direct evaluation of the integral in Eq.~(\ref{I}) is an intractable
task owing to the logarithm appearing in the integrand (a detailed analysis
of the difficulties arising in an analytical treatment of the mutual
information can be found in the PhD thesis of Fuchs \cite{Fuchs}). However,
applying Jensen's inequality \cite{Fuchs} it enables us to develop an upper
bound to the function $I(\vartheta)$ given by Eq.~(\ref{I}) and to its
weighted sum, the information gain~(\ref{Iav}). Jensen's inequality
involving a probability density function can be stated as follows
\cite{Rudin}: If $g$ is any real valued measurable function, $f$ is a
probability density function, and $\varphi$ is concave over the range of $g$,
then
\begin{equation}
\int{dn f(\mathbf n)\varphi(g(\mathbf n)) \leq \varphi\left(\int{dn
f(\mathbf n)g(\mathbf n)}\right)}\;. \label{Jensen}
\end{equation}
Let the concave function $\varphi$  be particularly the logarithm function
$\log_2 x$, and let functions $f$ and $g$ be equal to the fraction
${p(c_r,\vartheta|\mathbf n)}/{p(c_r,\vartheta)}$. As a result a complete
correspondence can be established between the integral within Eq.~(\ref{I})
and the left-hand side of Eq.~(\ref{Jensen}), entailing the upper bound
$J(\vartheta)$ to the function $I(\vartheta)$ as follows:
\begin{equation}
I(\vartheta) \leq \frac{3-\cos\vartheta}{12}\log_2\left( \int{dn \left(
\frac{p(c_r,\vartheta|\mathbf n)}{p(c_r,\vartheta)}\right)^2}\right)\equiv
J(\vartheta)\;. \label{J}
\end{equation}
Note that in contrast to the function $I(\vartheta)$, which can be computed
only numerically, its upper bound $J(\vartheta)$ can be given in analytic
terms. The respective curves of the function $I(\vartheta)$ and the function
$J(\vartheta)$ are plotted in Fig.~1 in the range $0\leq\vartheta\leq\pi$.

\begin{figure}[h]
\vspace*{-0.5cm} \centering \hspace*{-3cm}
\includegraphics[scale=1]{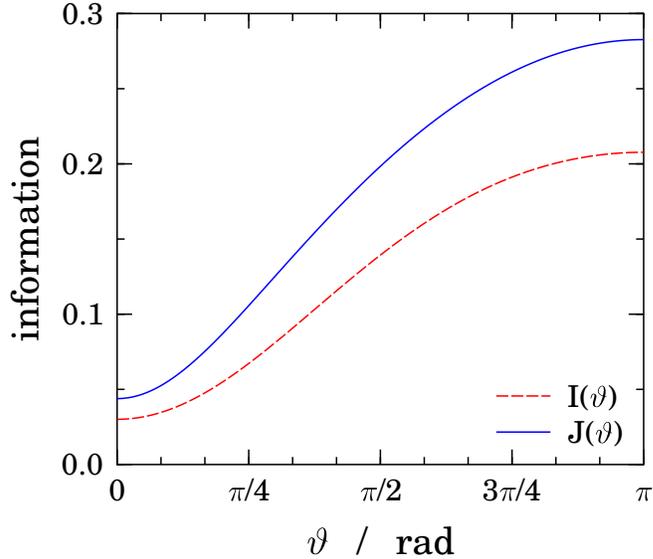}
\vspace{-1.5cm} \caption{The function $I$ and its upper bound $J$ plotted
against $\vartheta$ in the interval $0\leq\vartheta\leq\pi$.}
\end{figure}

After developing an upper bound to the information function $I(\vartheta)$
we wish to show that finding a global maximum of the function
\begin{equation}
J_{av}=\sum_{r=1}^M{c_rJ(\vartheta_r)} \label{Jav}
\end{equation}
in the feasible region (which region is defined in Subsection~(3.2)) will
serve as an upper bound to the global maximum of the information
gain~(\ref{Iav}) in the same feasible region, i.e., to the amount of
information which Bob can acquire by his best unentangled measurement.

To supply a proof, let us suppose the opposite, that is inside the feasible
region the maximum value of $I_{av}$ is bigger than the maximum value of
$J_{av}$. However, owing to the positive weights $c_r$ and the fact that
$I(\vartheta)\leq J(\vartheta)$, by the definitions~(\ref{Iav}, \ref{Jav})
$I_{av}\leq J_{av}$ at any point of the feasible region. Thus, by means of
this argument $I_{av}$ should also be upper bounded by $J_{av}$ at the very
point of its maximum inside the feasible region, which contradicts our
assumption, thereby completing the proof.

\subsection{Lagrange multipliers}
In this subsection, by the method of Lagrange multipliers we find via an
analytical treatment the value of the global maximum of $J_{av}$ in the
feasible region so as to provide an upper bound to the highest value of
$I_{av}$ in the feasible region (as stated in the preceding subsection),
achievable by an unentangled measurement. Thus we will obtain an upper bound
to the amount of information which Bob can gain about Alice's states by
carrying out unentangled measurements.

To this end, let us introduce the Lagrange multipliers $\lambda_1$ and
$\lambda_2$ which aim to account for the constraints~(\ref{newconstraints}).
Note that inequality constraints~(\ref{space}) will instead be taken into
account by restricting the domain of solutions. Then our task is to maximize
the Lagrangian $L$,
\begin{equation}
L=\sum_{r=1}^M{c_r J(\theta_r)}+\lambda_1\sum_{r=1}^M{c_r
\cos\theta_r}+\lambda_2\left(\sum_{r=1}^M{c_r}-4\right)\;. \label{L}
\end{equation}
Variations of $L$ with respect to $\theta_r$ and $c_r$ yield the following
two sets of equations,
\begin{equation}
\frac{\delta L}{\delta \theta_r}=0\;,\hspace{7mm} \frac{\delta L}{\delta
c_r}=0\;,\hspace{7mm} r = 1, \ldots, M\;, \label{setL}
\end{equation}
which can be solved for $\lambda_1$ and $\lambda_2$, and we obtain
\begin{align}
\lambda_1 &=\frac{1}{\sin\theta_r}\frac{d J(\theta_r)}{d
\theta_r}\;,  \nonumber\\
\lambda_2 &=-J(\theta_r)-\cot(\theta_r)\frac{d J(\theta_r)}{d \theta_r}\;,
\hspace{7mm} r = 1, \ldots, M\;. \label{lambda}
\end{align}
Let us define the function
\begin{equation}
h(\vartheta)=\frac{1}{\sin\vartheta}\frac{d J(\vartheta)}{d
\vartheta}\;.\label{h}
\end{equation}
Then the first equality within Eq.~(\ref{lambda}) becomes
$\lambda_1=h(\theta_r)$. Next our aim is to characterize $h(\vartheta)$
according to its monotonicity. Differentiating $h(\vartheta)$ with respect
to $\vartheta$ we obtain the explicit formula
\begin{equation}
\frac{d h(\vartheta)}{d\vartheta}= \frac{-16}{3\ln 2}
\frac{15-8\cos\vartheta+\cos2\vartheta}{\left(27-20\cos\vartheta+\cos
2\vartheta\right)^2}\frac{\sin\vartheta}{3-\cos\vartheta}\;, \label{deltah}
\end{equation}
which is negative in the range $0< \vartheta <\pi$ (and zero at
$\vartheta=0,\pi$) implying that $h(\vartheta)$ is a strictly decreasing
function in the interval $0< \vartheta <\pi$. Further, according to
Eq.~(\ref{lambda}), $h(\theta_r)$ must be equal to a yet undetermined
constant $\lambda_1$ for $r = 1, \ldots, M$ at a stationary point (which can
be either a point of local extremum or a saddle point) in the feasible
region. Thus the monotonicity of $h(\vartheta)$ implies that at a
stationary point in the feasible region all $\theta_r$ must be the same,
that is one single solution exists for the variables $\theta_r$,
\begin{equation}
\theta_r=\theta_{opt} \hspace{7mm} r = 1, \ldots, M\;. \label{optimum}
\end{equation}

In order to determine unambiguously $\theta_{opt}$ let us invoke
constraints~(\ref{newconstraints}), which allow us to write at the above
stationary point the following chain of equalities:
\begin{equation}
\sum_{r=1}^M{c_r
\cos\theta_r}=\sum_{r=1}^M{4\cos\theta_{opt}}=4\cos\theta_{opt}=0\;.
\label{chain}
\end{equation}
Hence the last equality provides us with the explicit solution
\begin{equation}
\theta_r=\theta_{opt}=\frac{\pi}{2}\;, \hspace{7mm} r = 1, \ldots, M
\label{solution}
\end{equation}
in the interval $[0,\pi]$, whereas the values of weights $c_r$ must satisfy
the condition $\sum_{r=1}^M{c_r=4}$ (i.e., they are in the feasible region).
Applying this solution~(\ref{solution}) and the corresponding condition we
may write at this stationary point for the value of $J_{av}$,
\begin{equation}
\max J_{av}=\sum_{r=1}^M{c_r J(\theta_{opt})}=4 J(\pi/2)=0.7935 \;
\mathrm{bits}\;. \label{maxJav}
\end{equation}

Now we wish to prove that the value of $\max J_{av}$ is a global maximum of
the function $J_{av}$ inside the feasible region. Further, it is sufficient
to show that it is a local maximum due to the single
solution~(\ref{solution}).

For this aim let us fix the values of $c_r$ in the feasible region, and
evaluate the Hessian matrix of $J_{av}(\theta_1,\dots,\theta_M)$ at the
point of the solution~(\ref{solution}). After differentiations we obtain the
Hessian as an $M\times M$ diagonal matrix whose $k$-th diagonal entry is
given by $-56c_k/(1521\ln 2)$. Since the weights $c_k$ are positive, the
Hessian matrix is negative definite implying that the
solution~(\ref{solution}) is a point of local maximum in an unrestricted
domain of $\theta_r$ and consequently it is in the (smaller) feasible region
as well.

This proves our proposition that $\max J_{av}=0.7935$ bits is a global
maximum, which can be attained by the function $J_{av}$ subject to the
constraints~(\ref{space}-\ref{newconstraints}). Combining this result with
the argument given in the previous subsection entails that the value
$0.7935$ bits necessarily upper bounds the information gain~(\ref{Iav})
attainable by an unentangled measurement on Alice's two antiparallel spins.

We found this upper bound by an analytical treatment, however by means of
numerical calculations we may arrive as well at the maximum information gain
$\max I_{av}$ attainable by an unentangled measurement if one replaces
$I\rightarrow J$ in Eqs.~(\ref{L}-\ref{h}). Owing to the logarithm in the
integrand~(\ref{I}) this really needs numerical integration. Numerics shows
that $h(\vartheta)$ will be a monotonic decreasing function in this case as
well, providing the same stationary point~(\ref{solution}) in the feasible
region for the information gain~$I_{av}$ as it was found before for its
upper bound $J_{av}$. In the present case, however, we obtain
\begin{equation}
\max I_{av}=\sum_{r=1}^M{c_r I(\theta_{opt})}=4 I(\pi/2)=0.557 \;
\mathrm{bits}\;. \label{maxIav}
\end{equation}

By applying the same arguments for $I_{av}$ as for its upper bound $J_{av}$
and by evaluating the Hessian matrix (which can be done this time only
numerically), we conclude that the solution~(\ref{solution}) is a point of
global maximum of $I_{av}$ in the feasible region (as for $J_{av}$), and
therefore we can assert that the maximum mutual information between Bob's
unentangled measurement and Alice's antiparallel spins is $\max
I_{av}=0.557$ bits.

In the next subsection we discuss the concrete form of the POVMs which
corresponds to the solution~(\ref{solution}), implying that the values $\max
I_{av}=0.557$ bits and $\max J_{av}=0.7935$ bits indeed correspond to a
realizable measurement.

\subsection{POVMs}

Our aim is to obtain those POVM elements $E_r$ within Eq.~(\ref{POVM}) which
produce Bob's best unentangled measurement. For this, we substitute the
solution~(\ref{solution}) into Eq.~(\ref{nr}) and as a result the mapped
states $|\tilde{\psi_r}\rangle$ in Eq.~(\ref{map}) become
\begin{equation}
|\tilde{\psi_r}\rangle=|\mathbf z\rangle\left(\frac{|\mathbf
z\rangle+|-\mathbf z\rangle}{\sqrt 2}\right)\equiv|B\rangle \;, \label{B}
\end{equation}
i.e., each of them turns out to be the same ($r$ independent) reference state
$|B\rangle$. Now, inverting the map~(\ref{map}) and defining the unit
vectors $\mathbf m_r$ through the arbitrary spatial rotations $\mathbf
m_r=R_r(\mathbf z)$ yield
\begin{equation}
|\psi_r\rangle=|\mathbf m_r\rangle\left(\frac{|\mathbf m_r\rangle+|-\mathbf
m_r\rangle}{\sqrt 2}\right) \;. \label{psir}
\end{equation}
Let us try with a minimal measurement, i.e., a measurement which has the
minimum number $M=4$ of POVM elements. This corresponds to a von Neumann
measurement satisfying the orthogonality requirement $E_r E_s = E_r
\delta_{rs}$. Consequently, $\langle \psi_r|\psi_s\rangle=\delta_{rs}$
implying $c_r=1,\; r=1,2,3,4$ (in accord with conditions in
Eqs.~(\ref{space}-\ref{newconstraints}) for $c_r$). Now we are left with
finding the angles $(\theta_r^m,\varphi_r^m)$ defining directions $\mathbf
m_r,\; r=1,2,3,4$, so as to completely define the POVM elements. If we choose
the angles $(\theta_r^m,\varphi_r^m)$ to be
\begin{equation}
(0,0) \hspace{5mm} (0,\pi) \hspace{5mm} (\pi,0) \hspace{5mm} (\pi,\pi)\;,
\end{equation}
it can be verified that the corresponding states $|\psi_r\rangle$ in
Eq.~(\ref{psir}) indeed constitue a legitimate POVM, $\sum_{r=1}^4
{|\psi_r\rangle\langle\psi_r|}=\mathbb I$. The measuring strategy described
by this unentangled POVM is in fact an LOCC measurement: Bob makes a von
Neumann measurement of Alice's first spin along an arbitrary direction (say
$\mathbf z$) and of Alice's second spin along an orthogonal direction.

On the other hand, Bagan et al.~\cite{Bagan2} found that for a pair of
antiparallel spins a measurement strategy which yields the maximal fidelity,
at the same time attains the value $0.8664$ bits of the mutual information .
The corresponding POVM measurement is a von Neumann type, described by
the projectors $E_r=|\psi_r\rangle\langle\psi_r|$ as follows \cite{Bagan1},
\begin{equation}
|\psi_r\rangle=\frac{\sqrt 3}{2}\frac{|\mathbf m_r\rangle|-\mathbf
m_r\rangle+|-\mathbf m_r\rangle|\mathbf m_r\rangle}{\sqrt 2} +
\frac{1}{2}|\psi^-\rangle \;,\hspace{6mm}r=1,2,3,4\;, \label{psirent}
\end{equation}
where the four unit vectors $\mathbf m_r$ are pointing to the vertices of a
tetrahedron inscribed in the unit sphere (given explicitly by
Ref.~\cite{Bagan1}) and $|\psi^-\rangle$ denotes the singlet state. All
four states in Eq.~(\ref{psirent}) are in fact entangled; thus these states
correspond to an entangled measurement. Incidentally, they ought to be
entangled owing to our analysis as well, providing to the best unentangled
measurement the upper bound $0.7935$ bits of mutual information (which is
smaller than $0.8664$ bits). Though it seems difficult to prove
analytically that the value $0.8664$ bits is the accessible information
corresponding to the most informative measurement on Alice's signal state,
we carried out extensive numerical calculations which support this
conjecture. Nevertheless, the value $0.8664$ bits definitely lower bounds
the mutual information attainable by an entangled measurement, and the value
$0.7935$ bits obtained in the preceding subsection upper bounds the mutual
information attainable by an unentangled measurement. Therefore, the nonzero
gap between the two bounds provides us with the proof that in general
optimal state discrimination cannot be achieved by an unentangled
measurement, if the performance of the state discrimination is quantified by
the mutual information.

\subsection{Parallel spins}

We may directly obtain results from our previous analysis for the case when
Alice uses two parallel spins to encode information. Actually, one needs to
flip the second spin $|-\mathbf n\rangle$ into $|\mathbf n\rangle$ in
Eq.~(\ref{mapprn}) which affects Eq.~(\ref{mappr}) as well, and then
substitute these modified formulas into the information gain~(\ref{Iav}).
However, the flip of the second spin is equivalent in effect to flip the
direction $\mathbf n_r\rightarrow-\mathbf n_r$ in Eq.~(\ref{mapprn}).
Especially, symmetry requires that the one-to-one correspondence between the
case of parallel and antiparallel spins is given by the change of variables
$\theta_r\rightarrow\pi-\theta_r$ in the formula for the information
gain~(\ref{Iav}). Taking into account the above mapping the
solution~(\ref{solution}) for antiparallel spins also holds true for
parallel spins. Thus the best unentangled measurement on parallel spins
(such as on antiparallel spins) is LOCC type, associated with
states~(\ref{psir}), providing the same mutual information $\max
I_{av}=0.557$ bits as in Section~(4.2) for two antiparallel spins.
Actually, this result can be seen from the outset if we recall that in the
case of LOCC protocols there is no difference between performing
measurements on parallel and antiparallel spins \cite{Massar}.

On the other hand, the optimal measurement of parallel spins due to Tarrach and
Vidal \cite{TV} is the one which is defined by the entangled states
\begin{equation}
|\psi_r\rangle=\frac{\sqrt 3}{2}|\mathbf m_r\rangle|\mathbf
m_r\rangle+\frac{1}{2}|\psi^-\rangle \;, \hspace{6mm}r=1,2,3,4
\label{psirupup}
\end{equation}
where $\mathbf m_r$ are pointing to the four corners of the tetrahedron,
as in the antiparallel situation, given by Ref.~\cite{Bagan1}. The information
gain of this optimal measurement is $\log_2 3-(2/3)\log_2 e=0.623$ bits as
given by Ref.~\cite{TV}. Thus in the parallel case as well the best measuring
strategy of Bob proves to be an entangled measurement.

\section{Discussion}

In summary, an analytical proof was presented that the accessible
information obtainable by an optimal measurement about a random space
direction $\mathbf n$ encoded in a pair of antiparallel spins cannot be
attained by an unentangled measurement. The information gain has been
quantified by the Shannon mutual information between the signal states and
the measurement outcomes, and by an unentangled measurement we mean that
each POVM operator is a tensor product.

We used a particular form of the mutual information, well-defined for a
continuous distribution of the signal states, and exploited its rotational
invariance. Then Jensen's inequality enabled us to upper bound the mutual
information attainable by an unentangled measurement. This upper bound has
been found by the Lagrange multipliers method. Explicitly, we obtained the
upper bound $0.7935$ bits of information for the best unentangled
measurement while the lower bound $0.8664$ bits of information corresponds
to the best entangled measurement.

We also made numerical calculations, which revealed that the maximum mutual
information which can be attained by an unentangled measurement is $0.557$
bits both for the cases of antiparallel and parallel spins, and in turn both
correspond to the same von Neumann type measurement apparatus. This entails
that interestingly for the case of antiparallel spins the optimal
measurement is about one and one-half times more effective than an
unentangled measurement, and for the case of parallel spins it is still more
effective but to a lesser degree, provided that the measure of success is
given in terms of the mutual information.

Let us make a comparison between the case of antiparallel spins analyzed in
this article, and the double-trine states of Refs.~(\cite{PW},
\cite{Wootters}) from the state distinguishability point of view. While on the
double-trine ensemble the best unentangled measurement was actually a global
measurement, for the antiparallel (and also for the parallel) spins the best
unentangled measurement was in turn an LOCC measurement (especially
individual von Neumann type). This fact may partially explain the large
difference obtained in the power of unentangled and entangled measurements
on antiparallel spins, and also would raise the possibility of finding a state
ensemble, where the power of unentangled measurement lies between the power
of entangled and the power of LOCC measurements.
\\
\\
\noindent {\Large\bf Acknowledgements}
\\
\\
\noindent I would like to thank Professor W.~K. Wootters
 for several discussions, which inspired me to work on this subject.
 This work was supported by the Grant \"Oveges from the National Office for
 Research and Technology.


\begin{thebibliography}{99}

\bibitem{Helstrom}
C.W. Helstrom, {\it Quantum Detection and Estimation Theory} (Academic
Press, 1976).

\bibitem{Fuchs}
C.A. Fuchs, {\it Distinguishability and Accessible Information in Quantum
Theory, PhD thesis}, (University of New Mexico, 1995) arXiv:
quant-ph/9601020.

\bibitem{Peres}
A. Peres, {\it Quantum Theory: Methods and Concepts} (Kluwer Academic,
Dordrecht, 1995).

\bibitem{Shor}
P.W. Shor, IBM J. Res. \& Dev. {\bf 48} (1), 115 (2004).

\bibitem{PW}
A. Peres and W.K. Wootters, Phys. Rev. Lett. {\bf 66}, 1119 (1991).

\bibitem{Decker}
T. Decker, ArXiv: quant-ph/0509122.

\bibitem{Bennett1} C.H. Bennett, D.P. DiVincenzo, C.A. Fuchs, T. Mor, E. Rains,
P.W. Shor, J.A. Smolin, and W.K. Wootters, Phys. Rev. A {\bf 59}, 1070
(1999).

\bibitem{Bennett2} C.H. Bennett, D.P. DiVincenzo, T. Mor, P.W. Shor, J.A. Smolin,
and B.M. Terhal, Phys. Rev. Lett. {\bf 82}, 5385 (1999).

\bibitem{DiVincenzo}
D.P. DiVincenzo, T. Mor, P.W. Shor, J.A. Smolin, and B.M. Terhal, Commun.
Math. Phys. {\bf 238}, 379 (2003).

\bibitem{Wootters}
W.K. Wootters, Int. J. Quant. Inf. {\bf 4}, 219 (2006).

\bibitem{GP}
N. Gisin and S. Popescu, Phys. Rev. Lett. {\bf 83}, 432 (1999).

\bibitem{Massar}
S. Massar, Phys. Rev. A {\bf 62}, 040101(R) (2000).

\bibitem{PS}
A. Peres and P.F. Scudo, Phys. Rev. Lett. {\bf 87}, 167901 (2001).

\bibitem{Bagan1}
E. Bagan, M. Baig, A. Brey, R. Mu\~{n}oz-Tapia, and R. Tarrach, Phys. Rev.
A {\bf 63}, 052309 (2001).

\bibitem{BM}
E. Bagan and R. Mu\~{n}oz-Tapia, Int. J. Quantum Inf. {\bf 4}, 5 (2006).

\bibitem{Jeffrey}
E.R. Jeffrey, J.B. Altepeter, M. Colci, and P.G. Kwiat, Phys. Rev. Lett.
{\bf 96}, 150503 (2006).

\bibitem{Bartlett1}
S.D. Bartlett, T. Rudolph, and R.W. Spekkens, ArXiv: quant-ph/0610030.

\bibitem{Davies}
E.B. Davies, IEEE Trans. Inform. Theory {\bf IT24}, 596 (1978).

\bibitem{TV}
R. Tarrach and G. Vidal, Phys. Rev. A {\bf 60}, R3339 (1999).

\bibitem{ATV}
A. Ac\'in, R. Tarrach, and G. Vidal, Phys. Rev A {\bf 61}, 062307 (2000).

\bibitem{Bartlett2}
S.D. Bartlett, T. Rudolph, and R.W. Spekkens, Phys. Rev A {\bf 70}, 032321
(2004).

\bibitem{Hobson}
A. Hobson, J. Stat. Phys. {\bf 1}, 383 (1969).

\bibitem{Rudin}
W. Rudin, {\it Real and Complex Analysis} (McGraw-Hill, New York, 1987).



\bibitem{Bagan2}
E. Bagan, M. Baig, A. Brey, R. Mu\~{n}oz-Tapia, and R. Tarrach, Phys. Rev.
Lett. {\bf 85} 5230 (2000).

\end{thebibliography}
\end{document}